\documentclass[lettersize,journal]{IEEEtran}
\usepackage{amsmath,amsfonts}
\usepackage{algorithmic}
\usepackage{algorithm}
\usepackage{array}
\usepackage[caption=false,font=normalsize,labelfont=sf,textfont=sf]{subfig}
\usepackage{textcomp}
\usepackage{stfloats}
\usepackage{url}
\usepackage{verbatim}
\usepackage{graphicx}
\usepackage{cite}
\usepackage{pifont}

\usepackage{multirow}
\usepackage{hyperref}
\begin{document}

\title{ASRRL-TTS: Agile Speaker Representation Reinforcement Learning for Text-to-Speech Speaker Adaptation}

\author{Ruibo Fu$^{\dag,*}$, \emph{Member, IEEE,}
  Xin Qi$^{\dag}$, \emph{Student Member, IEEE,}
  Zhengqi Wen, \emph{Member, IEEE,}
  Jianhua Tao, \emph{Senior Member, IEEE,}
  Tao Wang, \emph{Student Member, IEEE,}
  Chunyu Qiang, \emph{Student Member, IEEE,}
  Zhiyong Wang, \emph{Student Member, IEEE,}
  Yi Lu, \emph{Student Member, IEEE,}
  Xiaopeng Wang,  \emph{Student Member, IEEE,}
  Shuchen Shi, \emph{Student Member, IEEE,}
  Yukun Liu, \emph{Member, IEEE,}
  Xuefei Liu, \emph{Member, IEEE,}
  Shuai Zhang, \emph{Member, IEEE,}

\thanks{Ruibo Fu, Xin Qi, Zhengqi Wen, Tao Wang, Yi Lu, Zhiyong Wang, Xiaopeng Wang and Xuefei Liu are with Institute of Automation, Chinese Academy of Sciences, Beijing, China. }

\thanks{ Xin Qi, Yi Lu, Zhiyong Wang,, Xiaopeng Wang, Yukun Liu are with School of Artificial Intelligence, University of Chinese Academy of Sciences, Beijing, China.}
\thanks{Jianhua Tao and Shuai Zhang are with Department of Automation, Tsinghua University, Beijing, China. Jianhua Tao is also with Beijing National Research Center for Information Science and Technology, Tsinghua University, Beijing, China.}
\thanks{ Chunyu Qiang is with Kuaishou Technology Co., Ltd., Beijing, China, and also with Tianjin University, Tianjin, China.}
\thanks{ Shuchen Shi is with Shanghai Polytechnic University, China.}

\thanks{Ruibo Fu and Xin Qi contribute equally to this work. Ruibo Fu is the corresponding author. (e-mail: ruibo.fu@nlpr.ia.ac.cn)}}

\markboth{Journal of \LaTeX\ Class Files,~Vol.~14, No.~8, August~2021}%
{Shell \MakeLowercase{\textit{et al.}}: A Sample Article Using IEEEtran.cls for IEEE Journals}


\maketitle

\begin{abstract}
Speaker adaptation, which involves cloning voices from unseen speakers in the Text-to-Speech(TTS) task, has garnered significant interest due to its numerous applications in multi-media fields. Despite recent advancements, existing methods often struggle with inadequate speaker representation accuracy and overfitting, particularly in limited reference speeches scenarios. To address these challenges, we propose an Agile Speaker Representation Reinforcement Learning (ASRRL) strategy to enhance speaker similarity in speaker adaptation tasks. ASRRL is the first work to apply reinforcement learning (RL) to improve the modeling accuracy of speaker embeddings in speaker adaptation, addressing the challenge of decoupling voice content and timbre. Our approach introduces two action strategies tailored to different reference speeches scenarios. In the single-sentence (SS) scenario, a knowledge-oriented optimal routine searching RL method is employed to expedite the exploration and retrieval of refinement information on the fringe of speaker representations. In the few-sentence (FS) scenario, we utilize a dynamic RL method to adaptively fuse reference speeches, enhancing the robustness and accuracy of speaker modeling. To achieve optimal results in the target domain, a multi-scale fusion scoring mechanism based reward model that evaluates speaker similarity, speech quality, and intelligibility across three dimensions is proposed, ensuring that improvements in speaker similarity do not compromise speech quality or intelligibility. The experimental results on the LibriTTS and VCTK datasets within mainstream TTS frameworks demonstrate the extensibility and generalization capabilities of the proposed ASRRL method. The results indicate that the ASRRL method significantly outperforms traditional fine-tuning approaches, achieving higher speaker similarity and better overall speech quality with limited reference speeches. The audio demo is available at \url{https://7xin.github.io/ASRRL/}.
\end{abstract}

\begin{IEEEkeywords}
speaker adaptation, reinforcement learning, TTS, action design, reward model
\end{IEEEkeywords}

\section{Introduction}

\IEEEPARstart{I}{n} recent years, there has been remarkable progress in the development of Text-to-Speech (TTS) technology\cite{wang17n_interspeech,kim2021conditional,ijcai2022p577,chen20r_interspeech,gibiansky2017deep,kharitonov2023speak,betker2023better,jiang2023mega}, where the overall quality of the generated speech is close to or even surpasses that of real human speech. Among these advancements, speaker adaptation with few target reference speeches has also attracted extensive attention from academia and industry. For example, in entertainment scenes such as movies, dubbing, and broadcasting, many application scenarios generate personalized speech with a few number of target reference speeches. However, the current technology is still limited by the modeling accuracy of speaker representation and data constraints of reference speeches.

The existing technical approaches to the TTS speaker adaptation task primarily focus on two aspects: speaker representations and model fine-tuning. On the one hand, the aim is to train a more generalized TTS model and achieve accurate modeling of the target speaker by building the speaker encoder, thus enabling control over the target timbre. The current method involves extracting speaker representations from few reference speeches by using a speaker encoder, which are then conditionally input into the acoustic model for speaker adaptation. Tacotron 2\cite{8461368} directly obtains embeddings for unseen speakers through a speaker encoder trained with the Generalized End-to-End (GE2E) loss\cite{8462665}. Similar work includes YourTTS\cite{casanova2022yourtts} and MegaTTS\cite{jiang2023mega} which replace the input speaker representation with the speaker encoder's results. Large language models, such as VALL-E\cite{wang2023neural}, also adopt a similar method to extract discrete speaker representations. However, these methods often lack precision in extracting speaker information, leading to noticeable differences in timbre and perception compared to the target speaker\cite{DBLP:journals/corr/abs-2204-00990,lee23f_interspeech,casanova2021sc}. On the other hand, a series of transfer learning methods are used to fine-tune the entire TTS model\cite{kons19_interspeech,NEURIPS2018_4559912e}. These approaches have resulted in higher speaker similarity compared to the zero-shot scenario. However, with limited reference speeches, TTS models face over-fitting issues. Besides, the 
 low speech data quality would also lead to a decline in the performance of the synthestic speech \cite{zhou22d_interspeech}.

To alleviate the over-fitting problem, a technique that freezes the model weights and utilizes model plugins like LoRA\cite{hu2021lora} has been used to record new knowledge without having to alter the model itself. In these works, Hsieh et al.\cite{hsieh2022adapter} inserted LoRA within the attention module and, in conjunction with BitFit\cite{zaken2021bitfit} and other adapters\cite{houlsby2019parameter}, facilitated the model's learning of new speakers through a plug-and-play method. This method helps prevent over-fitting issues caused by direct fine-tuning. However, making changes to the model structure requires expert experience and comes with a higher design cost.

In this paper, inspired by the multiple applications of reinforcement learning in current Artificial Intelligence Generated Content (AIGC) technology\cite{kaufmann2023survey}, we propose the Agile Speaker Representation Reinforcement Learning (ASRRL) method, which is target domain refined similarity enhancement with agile speaker representation reinforcement learning in TTS speaker adaptation. In terms of model architecture, reinforcement learning is used to improve the modeling ability of speaker embedding, effectively enhancing the performance of speech synthesis without fine-tuning the whole TTS model, ensuring that each generated speech can achieve relatively optimal performance. For the two typical scenarios of the TTS speaker adaptation task, we set different state initialization methods and action design methods according to the situation. Furthermore, by designing a comprehensive reward model for speech synthesis considering multiple dimensions of similarity, quality, and intelligibility, the optimized search path of the RL agent is further refined, improving efficiency accordingly. The experiment demonstrates the extensibility and generalization of our proposed method through the LibriTTS\cite{zen19_interspeech} dataset and the VCTK\cite{Veaux2017CSTRVC} dataset in mainstream VITS \cite{kim2021conditional} model based on the VAE-flow framework and Grad-TTS\cite{ho2020denoising} model based on the diffusion framework. The proposed ASRRL method effectively improves the performance of speaker adaptation with few samples, surpassing the performance of fine-tuning the entire TTS model.

The main contributions of this paper are as follows:
\begin{itemize}
\item To the best of our knowledge, ASRRL is the first work to apply reinforcement learning to enhance the modeling accuracy of speaker representation in speaker adaptation. This method effectively addresses the problem of speaker representation modeling accuracy caused by the difficulty of completely decoupling the voice content and timbre in speaker modeling, further improving the performance of speech synthesis for specific domains.
\item Two distinct reinforcement learning action strategies are designed based on the number of reference speeches samples. In the Single-Sentence(SS) scenario, prior knowledge is employed as guidance to expedite the refinement on the fringe of speaker representation. In the Few-Sentence(FS) scenario, a dynamic fusion reinforcement learning method is utilized to merge a few reference speeches.
\item A score fusion mechanism is designed to ensure an improvement in the similarity of synthetic speech while avoiding catastrophic declines in quality and intelligibility, thereby ensuring the consistency of speech performance and human perception.
\end{itemize}

\section{Related Work}
\subsection{Speaker Adaptation}
Speaker adaptation has always been a hot topic, with many related studies or applications \cite{ping2017deep,casanova2022yourtts}.
The common methods mainly include two approaches: fine-tuning and speaker encoding.
In the line of fine-tuning, related works \cite{chen2018sample, wang2020spoken, chen2021adaspeech} mainly focus on improving the efficiency of fine-tuning, enabling the model to learn new speakers using less data.
In the line of speaker encoding, neural networks are used to extract speaker embedding from reference speech. 
This route demonstrates good generalization, as providing speech from any speaker ultimately generates corresponding constraints. 
The Glow-WaveGan2\cite{lei22_interspeech} utilizes a neural encoder to generate speaker feature constraints for TTS. Simultaneously, it employs a joint training encoder based on speaker classification to learn speaker representations. SC-GlowTTS\cite{casanova2021sc} employs a three-layer LSTM as a speaker encoder. They use local sensitive attention to embed speaker representation connected to the input of the attention module.

Currently, the development of large language models is progressing rapidly. In the field of speech models based on large language model, their powerful generalization and fitting capabilities have quickly gained widespread attention. large language models use a speech codec to extract speaker representations from reference speech as discrete tokens. Microsoft has introduced their large language model VALL-E\cite{wang2023neural}, primarily structured using the Transformer\cite{vaswani2017attention}, encoding speakers with an Audio Codec Encoder\cite{defossez2022high}. It can synthesize the speech of a target person with just a 3-second speech clip. The subsequent release, VALLE-X\cite{zhang2023speak}, expands its capabilities to cross-language scenarios while maintaining the encoder structure for speaker representation. Google has also presented their Spear-TTS\cite{kharitonov2023speak}, training a SoundStream neural codec\cite{9625818} to reconstruct speech, encoding certain acoustic details as discrete acoustic tokens. Similarly, this model can synthesize speech for an unseen speaker based on a 3-second speech clip without the need for additional speaker representation. In Tortoise TTS\cite{betker2023better} work, speech segments are transformed into Mel spectrograms, which serve as conditional input to the autoregressive generator and the diffusion\cite{ho2020denoising}decoder. These methods, despite differences in technical details, essentially synthesize speech from unseen speakers by modeling speaker representations. Therefore, improving their accuracy is crucial for enhancing the similarity in speaker adaptation. 

\subsection{Reinforcement learning}

Reinforcement learning(RL) has demonstrated its powerful capabilities over time. For example, DQN (Deep Q-Network) \cite{mnih2013playing}, a reinforcement learning method for playing games, surpassed human performance in three different games.

Currently, many works are leveraging RL to optimize the performance of generative models.
In the field of music generation, RL-Duet\cite{jiang2020rl} interprets human-inputted musical scores as the current state of the environment, providing it to the agent. The agent's output actions represent newly generated musical notes. Through a reward model evaluating multiple aspects, the system iterates to output the best results. Similarly, in the domain of natural language processing (NLP), Code-RL\cite{le2022coderl} explores the task of code generation. In this case, the language model tasked with generating code serves as the agent, and another evaluation network is introduced to predict the correctness of the generated code, providing feedback to the agent in the iterative process.

Since the advent of large language models, RL methods based on human feedback have come into the spotlight.
InstructGPT\cite{ouyang2022training} uses manually annotated data to train a reward model, guiding RL to fine-tune pre-trained large language models, aiming to generate results that align with human preferences. In addition to this, researchers have proposed the Tempera\cite{zhang2022tempera} model. Starting with adjusting prompts for input to large language models, it divides the input into different parts, designs a discrete action space to make changes to the input, and gradually finds the optimal prompt through continuous exploration in RL.

Despite the rapid development of RL methods for fine-tuning language models, their application in TTS lags behind due to several challenges. These challenges include how to integrate TTS with RL and how to design appropriate reward signals. Therefore, we propose ASSRL to explore the potential of RL applications in TTS by enhancing the similarity in speaker adaptation.

\begin{figure*}[t]
  \centering
  \includegraphics[width=\linewidth]{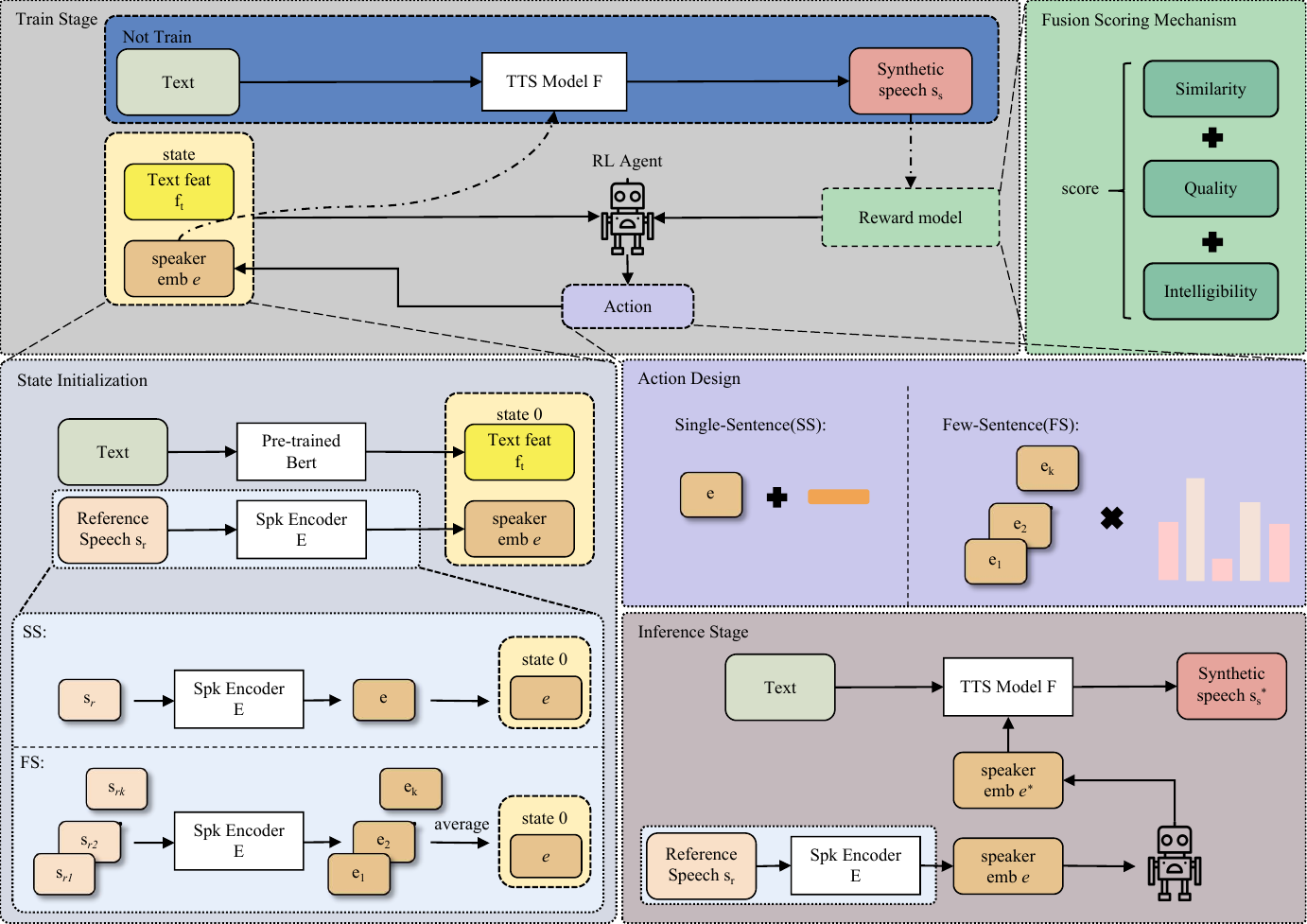}
  \caption{
The framework of ASSRL: 
(1)Extract the speaker representation embedding $e$ from the reference speech $s_r$. Additionally, extract text features $f_t$ using a pre-trained Bert model. These text features and speaker embeddings form the state representation for RL. A pre-trained TTS model $F$ synthesizes the text features and speaker embeddings into speech $s_s$. The reward model then evaluates the current generation results using a fusion scoring mechanism. Based on the scenario, the RL agent selects different actions to alter the speaker embedding information $e$ and obtain the next state. 
(2)When there are a few reference speeches, the initial state's speaker embedding part is represented by the mean of the speaker embeddings extracted by the encoder. 
(3)During inference, the trained policy is utilized to modify the speaker embedding information $e$, resulting in an enhanced speaker similarity denoted by $e^*$. 
  }
  \label{overview}
\end{figure*}

\section{Method}

This section introduces ASSRL, an optimization method for speaker adaptation. ASSRL enhances the similarity of synthesized speech for previously unseen speakers by using RL to finely tune speaker representations without altering the underlying model.

It will mainly be introduced from three aspects: (1) the prior state representation, which is designed to integrate RL with the speaker adaptation process; (2) the fusion scoring mechanism, which comprehensively evaluates the optimization process to ensure enhanced similarity without compromising generation quality; and (3) the action design for scenarios with one or a few reference speeches.

\subsection{The framework of ASSRL}
The overall framework of ASSRL is illustrated in Figure 1.
During training, the data input consists of text and reference speech.
The pre-trained BERT model extracts text features \( f_t \), and the speaker encoder extracts speaker embeddings \( e \). \( f_t \) and \( e \) form the prior state (State Initialization in the figure \ref{overview}).
When there are multiple reference speeches, the initial state is represented as the average fusion of the speaker embeddings.
Next, \( e \) and the text are processed by the TTS model \( F \) to synthesize speech \( s_s \). The synthesized speech \( s_s \) is then evaluated using a fusion scoring mechanism (Fusion Scoring Mechanism in the figure \ref{overview}).
Depending on the number of reference speeches, two different actions are taken (Action Design in the figure \ref{overview}): (1) In the SS scenario, a sensitive adjustment is made to each dimension of \( e \). (2) In the FS scenario, the fusion ratio of the speaker embeddings is adjusted.
During inference, the speaker embedding \( e \) is adjusted using the trained strategy to obtain a more accurate similarity embedding \( e^* \). Then, \( e^* \) and the text are used to generate speech, resulting in a synthesized output \( s_s^* \) with higher similarity.

\subsection{Deep generative models}

Our method is applicable to the current mainstream speech synthesis architectures, including VAE-Flow-GAN and Diffusion structures. For the former, we use VITS \cite{kim2021conditional} as a representative model, which takes text as input and models feature distribution for sampling. For the latter, we use the Grad-TTS \cite{popov2021grad} model as a representative, which decodes features into speech using a diffusion model.

To initially model the speaker representation of unseen speakers, we use a speaker encoder instead of distinguishing speakers through an embedding layer based on speaker IDs. The structure of the speaker encoder is a residual network.

\subsection{Agile Speaker Representation Reinforcement Learning}

\begin{figure}[h]
  \centering
  \includegraphics[width=\linewidth]{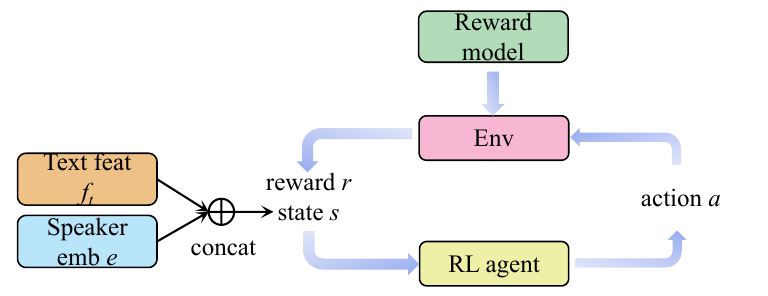}
  \caption{The process of Agile Speaker Representation Reinforcement Learning}
  \label{rl}
\end{figure}

Adjustments to speaker representation are made using an RL method, defining this process as a Markov decision process. The RL agent employs a transformer-based network architecture and the Proximal Policy Optimization (PPO) algorithm, one of the best-performing algorithms currently available. 
As shown in Figure \ref{rl}, the initial state \( s = (f_t, e) \) comprises the text features \( f_t \), extracted using a pre-trained model, and the speaker embedding \( e \), derived from the reference speech. 
Incorporating text features helps the RL agent better provide the optimal speaker representation for the current text situation due to the limited content generated for specific text compared to the entire model text space.
At each time step \( n \), the RL agent samples a series of actions \( a \), representing adjustments to the speaker's expression. 
At this stage, the TTS model and the reward model are encapsulated within an environment class (Env). The Env is responsible for receiving action information and updating the state, while the reward model calculates the reward \( r \) for the current action. 
The transition process can be defined as \( s_n = (f_{t_n}, e_n) \rightarrow s_{n+1} = (f_{t_{n+1}}, e_{n+1}) \).

In the SS scenario, we set the number of RL exploration steps to three. This decision is based on empirical judgment, as it is expected that the optimal speaker representation \( e^* \) and the speaker embedding \( e \) obtained from the encoder should not differ significantly. A longer exploration path may result in a greater discrepancy between the adjusted representation and the target speaker. This hypothesis was confirmed in subsequent experiments. In the FS scenario, where the objective is to adjust the fusion ratio of multiple speaker embeddings, the number of exploration steps is set to one.

\subsubsection{Prior-Guided State Representation}
The RL framework provides great flexibility in representing states. In our work, states are represented with a fixed length for ease of transition between states. During the experiments, we found that synthesizing speech with the same speaker embeddings but different texts results in varying degrees of speaker similarity. In the ablation experiment, we also verified that the addition of text features has a significant impact on the experimental results. Therefore, we introduce the prior knowledge of text feature \( f_t \) into the state. Finally, the separator \( <sep> \) and speaker embedding \( e \) are added together to form the state representation in RL. In the FS scenario, the initial state of speaker embedding \( e \) is represented by the mean of speaker embeddings extracted from a few reference speeches.

\begin{figure}[h]
  \centering
  \includegraphics[width=\linewidth]{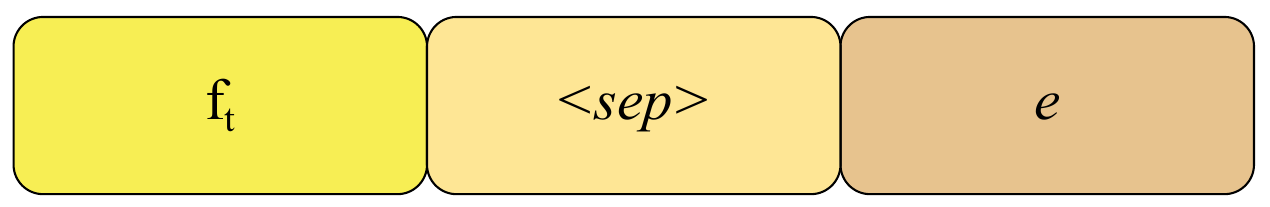}
  \caption{State Representation. Text features $f_t$. Speaker representation embedding $e$. Separator $<sep>$.}
  \label{state}
\end{figure}

\subsubsection{Two-Scenario Action Design}
\begin{figure}[h]
  \centering
  \includegraphics[width=\linewidth]{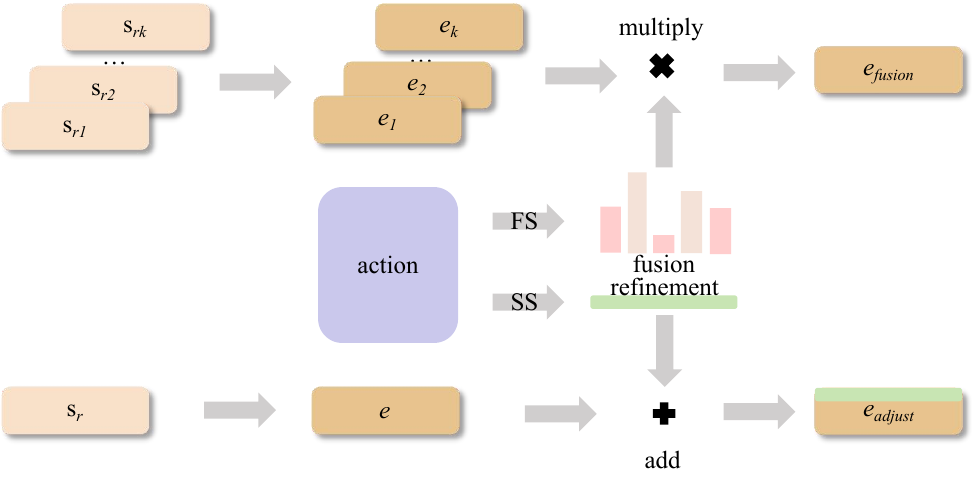}
  \caption{Two-Scenario Action Design. In the SS scenario, the output action is a refinement with the same dimensions as $e$. In the FS scenario, the output action is the fusion ratio for a few reference speeches.}
  \label{action}
\end{figure}

The speaker representation should be adjusted slightly and flexibly. At the same time, excessive changes will cause the synthesis results to be uncontrollable. Therefore, continuous actions need to be defined and limited to a given action space. Action designs are envisioned for two scenarios, as illustrated in Figure \ref{action}.

\textbf{Few-Sentence (FS):} When dealing with an unseen speaker with a few reference speeches \( s_{r1}, s_{r2}, \ldots, s_{rk} \) (\( k > 1 \)), speaker representations for all or some of these speeches are extracted as \( e_1, e_2, \ldots, e_k \) (\( k > 1 \)). The output of the action by the RL agent is considered as adjusting the fusion ratio for these speaker representations, with dimensions equal to \( k \). After adjustment, the fused speaker representation \( e_{fusion} \) is obtained. The transition between states is achieved by changing the proportion of speaker representation. Through continuous adjustments and evaluations, the RL agent discovers the optimal ratio in its ongoing exploration process. As these speaker representations are obtained from different reference speeches, they encompass timbres corresponding to a greater variety of phonemes. Additionally, since these speaker representations are directly provided by the encoder, their effectiveness in synthesizing valid speech is ensured. Adjusting the proportions between them can enhance similarity while maintaining greater stability.

\textbf{Single-Sentence(SS):} 
When the unseen speaker has only one reference speech \( s_r \), the speaker representation \( e \) for that sentence is extracted, and the actions are designed as continuous variables of the same dimension as \( e \). Due to the difficulty in quickly finding the optimal result with coarse-grained adjustments, restrictions were applied to the action space, and the output of the reinforcement learning action network was scaled by a factor of 1000. The actions are interpreted as a refinement to \( e \). In subsequent experiments, we tested the hyperparameter of the scaling factor of the action to find a suitable value.

\subsubsection{Multi-Dimensional Fusion Reward Design}
In the RL process, the reward model is responsible for evaluating the quality of each action at each time step, with higher reward values indicating better performance. In the current state \( s \), the pre-trained voiceprint analysis model implemented in \cite{X-vectors} is first used to score the speech synthesis results. This model provides a similarity score \( sc_{sim} \) for the generated result relative to the reference speech, ranging from 0 to 1, where higher scores indicate better similarity. If rewards are solely based on similarity scores, RL strategies often tend to pursue optimization in a single dimension, leading to a decrease in other aspects such as speech quality and intelligibility.

Therefore, an additional pre-trained model \cite{saeki22c_interspeech} was utilized to obtain the Mean Opinion Score (MOS) for speech quality, denoted as \( sc_{mos} \). Additionally, a pre-trained speech recognition model \cite{grosman2021xlsr53-large-english} for intelligibility scoring \( sc_{\text{in}} \) is incorporated, which evaluates the intelligibility between the semantic content of the speech and its corresponding text by calculating word error rates. The range of \( sc_{\text{in}} \) is from 0 to 1, with smaller values indicating lower error rates. The final result can be expressed by the following formula:

\begin{align}
    sc = sc_{sim}+\lambda_1 sc_{mos}-\lambda_2 sc_{in}
\end{align}%

Due to the range of MOS scores from 0 to 5, it is necessary to divide the MOS score by 5 to unify the scale between multiple scores. \( \lambda_1 \) is set to 0.5, and \( \lambda_2 \) is set to 0.1. The setting of these two hyperparameters was explored in the subsequent experimental section.

After scoring the state \( s \), the reward \( r \) for each time step is calculated as the difference between the current state's score \( sc_n \) and the previous state's score \( sc_{n-1} \). This can be expressed by the following formula:
\begin{align}
    r_n = sc_n(f_{t\textbf{n}},e_n)-sc_{n-1}(f_{t\textbf{n-1}},e_{n-1})
\end{align}

If the action at the current time step is positive optimization, then its state score should be greater than the previous state, and the reward calculation should be a positive value.

This method of reward calculation, which incorporates multidimensional scores, is more comprehensive compared to the traditional speech synthesis approach that uses reconstruction loss as a single objective. It considers multiple aspects and achieves better alignment between the generated audio quality and human perception to a certain extent.

\section{Experiment and Analysis}
\subsection{Experimental Setup}
\subsubsection{Dataset}
We used the LibriTTS\cite{zen19_interspeech} dataset and the VCTK\cite{Veaux2017CSTRVC} dataset. The LibriTTS-clean and LibriTTS-other subsets were used separately since we wanted to verify the robustness of our method in the presence of noise using the LibriTTS-other dataset.

The VCTK Corpus includes around 44 hours of speech data uttered by 110 English speakers with various accents. Each speaker reads out about 400 sentences, which were selected from a newspaper, the rainbow passage, and an elicitation paragraph used for the speech accent archive. We resampled all speeches to 16 kHz.

LibriTTS is a multi-speaker English corpus of approximately 585 hours of read English speech at 24kHz sampling rate. It is derived from the original materials (mp3 audio files from LibriVox and text files from Project Gutenberg) of the LibriSpeech corpus. We resampled all speeches to 16 kHz. 

LibriTTS-clean, LibriTTS-other, and VCTK are used as three separate datasets, with each synthesis model and encoder trained individually on one of these datasets. Ten speakers are randomly selected from the VCTK dataset, along with the test-clean and test-other datasets, to evaluate our method on unseen speakers.

\subsubsection{Task}
We trained the current mainstream VITS \cite{kim2021conditional} model based on the VAE-flow framework and the Grad-TTS \cite{ho2020denoising} model based on the diffusion framework to verify the effectiveness of the proposed method in enhancing speaker similarity in SS and FS scenarios. Subjective and objective experiments were conducted to evaluate speech quality, intelligibility, and speaker similarity. Each evaluation result was obtained by averaging the scores of 100 speeches. The results were compared with those of baseline methods and their fine-tuned versions. Additionally, variations in similarity results between the proposed method and fine-tuning were compared as the number of reference speech samples increased. Ablation experiments were performed to validate the effectiveness of the fusion scoring mechanism and state representation design. Finally, experiments were conducted to validate the rationality of hyperparameter settings. Each model was trained on a single NVIDIA GeForce RTX 3090 GPU on each dataset for one day.

\subsubsection{Metrics}
In terms of objective evaluation, pre-trained models were employed for assessment. The MOS rating model, specifically speech MOS \cite{saeki22c_interspeech}, was used to assign scores to the speech quality of the experimental results, with a scoring range of 0 to 5, where higher scores indicate better quality. We used a speech recognition model \cite{grosman2021xlsr53-large-english} to recognize semantic content and calculate the word error rate (WER) between the speech and the text to measure speech intelligibility. The final results range from 0 to 1, with lower values indicating better intelligibility. Additionally, a voiceprint analysis model \cite{X-vectors} generated similarity scores relative to the reference speech, with a scoring range of 0 to 1, where higher scores are desirable.
In the subjective evaluation, we invited 25 paid human evaluators to assess similarity, speech quality, and intelligibility. Prior to the evaluation, they received appropriate training to ensure that the scoring scale is consistent with the objective evaluation.

\subsection{Subjective evaluation}
\subsubsection{Single-Sentence scenario(SS)}

We will include the results of ground truth (GT) in the table for reference. Our method will be validated using two models and three datasets. \( vits \) and \( diffusion \) represent the results of direct synthesis. \( vits-ft\) and \( diffusion-ft \) represent the results of fine-tuning for 2000 steps using a reference speech sentence. \( vits-rl \) and \( diffusion-rl \) represent the results of using our proposed method.
\begin{table*}[]
\centering
\caption{SUBJECTIVE RESULT IN SS SCENARIO}
\label{tab:my-table}
\begin{tabular}{l|ccc|ccc|ccc}
\hline
Subjective-SS & \multicolumn{3}{c|}{libritts-clean}                                 & \multicolumn{3}{c|}{libritts-other}                                 & \multicolumn{3}{c}{vctk}                                            \\ \hline
Model      & \multicolumn{1}{c|}{sim $\uparrow$} & \multicolumn{1}{c|}{mos $\uparrow$} & intell ↓       & \multicolumn{1}{c|}{sim $\uparrow$} & \multicolumn{1}{c|}{mos $\uparrow$} & intell ↓       & \multicolumn{1}{c|}{sim $\uparrow$} & \multicolumn{1}{c|}{mos $\uparrow$} & intell ↓       \\ \hline
GT         & 0.83                     & 4.26                     & 0.15          & 0.67                     & 4.62                     & 0.19          & 0.85                     & 4.44                     & 0.17          \\ \hline
vits       & 0.42                     & 4.12                     & 0.25          & 0.34                     & 4.12                     & 0.23          & 0.49                     & 4.25                     & 0.23          \\
vits-ft    & 0.44                     & 4.03                     & 0.30          & 0.35                     & 4.03                     & 0.23          & 0.50                     & 4.15                     & 0.29          \\
vits-rl    & \textbf{0.56}            & \textbf{4.24}            & \textbf{0.23} & \textbf{0.44}            & \textbf{4.42}            & \textbf{0.21} & \textbf{0.61}            & \textbf{4.38}            & \textbf{0.19} \\ \hline
diffusion    & 0.45                     & 4.23                     & 0.33          & 0.40                     & 4.35                     & 0.25          & 0.44                     & 4.23                     & 0.31          \\
diffusion-ft & 0.44                     & 4.14                     & 0.35          & 0.38                     & 4.18                     & 0.25          & 0.46                     & 4.16                     & 0.34          \\
diffusion-rl & \textbf{0.57}            & \textbf{4.25}            & \textbf{0.29} & \textbf{0.45}            & \textbf{4.39}            & \textbf{0.23} & \textbf{0.63}            & \textbf{4.33}            & \textbf{0.27} \\ \hline
\end{tabular}
\end{table*}

The subjective experimental results indicate that our proposed method has improved speaker similarity, speech quality, and intelligibility in the SS scenario.

The similarity can be improved by up to 0.19.

\subsubsection{Few-Sentence scenario(FS)}
We include the evaluation results of the ground truth (GT) in the table for reference. We validate our method using two models across three datasets. In the FS scenario, we set the number of reference speeches samples to three sentences. \( vits \) and \( diffusion \) are the results synthesized using the average speaker information extracted from the reference speeches. \( vits-ft \) and \( diffusion-ft \) are the results after fine-tuning for 2000 steps with three reference speech sentences. \( vits-rl-3 \) and \( diffusion-rl-3 \) are the results using our proposed method with the action strategy in the FS scenario.

\begin{table*}[]
\centering
\caption{SUBJECTIVE RESULT IN FS SCENARIO}
\label{Sub-ms}
\begin{tabular}{l|ccc|ccc|ccc}
\hline
\multicolumn{1}{c|}{Subjective-FS} & \multicolumn{3}{c|}{libritts-clean}                                     & \multicolumn{3}{c|}{libritts-other}                                     & \multicolumn{3}{c}{vctk}                                                \\ \hline
\multicolumn{1}{c|}{Model}         & \multicolumn{1}{c|}{sim ↑} & \multicolumn{1}{c|}{mos ↑} & intell ↓      & \multicolumn{1}{c|}{sim ↑} & \multicolumn{1}{c|}{mos ↑} & intell ↓      & \multicolumn{1}{c|}{sim ↑} & \multicolumn{1}{c|}{mos ↑} & intell ↓      \\ \hline
GT                                 & 0.69                       & 4.28                       & 0.25          & 0.83                       & 4.62                       & 0.18          & 0.82                       & 4.49                       & 0.11          \\ \hline
vits                               & 0.51                       & 4.08                       & 0.26          & 0.43                       & 4.24                       & 0.27          & 0.59                       & 4.25                       & 0.28          \\
vits-ft                            & 0.46                       & 4.02                       & 0.27          & 0.49                       & 4.13                       & 0.24          & 0.59                       & 4.29                       & 0.32          \\
vits-rl-3                          & \textbf{0.61}              & \textbf{4.24}              & \textbf{0.14} & \textbf{0.58}              & \textbf{4.46}              & \textbf{0.20} & \textbf{0.67}              & \textbf{4.42}              & \textbf{0.18} \\ \hline
diffusion                          & 0.52                       & 4.14                       & 0.34          & 0.44                       & 4.18                       & 0.25          & 0.59                       & 4.21                       & 0.34          \\
diffusion-ft                       & 0.46                       & 4.16                       & 0.43          & 0.47                       & 4.37                       & 0.28          & 0.56                       & 4.15                       & 0.25          \\
diffusion-rl-3                     & \textbf{0.55}              & \textbf{4.28}              & \textbf{0.23} & \textbf{0.55}              & \textbf{4.53}              & \textbf{0.25} & \textbf{0.66}              & \textbf{4.48}              & \textbf{0.25} \\ \hline
\end{tabular}
\end{table*}

The experimental results indicate that our method performs better than direct fine-tuning across two models and three datasets. While enhancing speaker similarity, it also slightly improves speech quality and intelligibility.

\subsection{Objective evaluation}
\subsubsection{Single-Sentence scenario(SS)}
In the SS scenario, we evaluated the experimental results across two models and three datasets. Pre-trained models were used to assess the objective metrics of the generated results. The table includes the ground truth (GT) results for reference, with the evaluations of similarity and speech quality accurate to five decimal places.

\begin{table*}[]
\centering
\caption{OBJECTIVE RESULT IN SS SCENARIO}
\label{Obj-ss}
\begin{tabular}{l|ccc|ccc|ccc}
\hline
\multicolumn{1}{c|}{Objective-SS} & \multicolumn{3}{c|}{libritts-clean}                                     & \multicolumn{3}{c|}{libritts-other}                                     & \multicolumn{3}{c}{vctk}                                                \\ \hline
\multicolumn{1}{c|}{Model}        & \multicolumn{1}{c|}{sim ↑} & \multicolumn{1}{c|}{mos ↑} & intell ↓      & \multicolumn{1}{c|}{sim ↑} & \multicolumn{1}{c|}{mos ↑} & intell ↓      & \multicolumn{1}{c|}{sim ↑} & \multicolumn{1}{c|}{mos ↑} & intell ↓      \\ \hline
GT                                & 0.60734                    & 4.51352                    & 0.17          & 0.58833                    & 4.41131                    & 0.11          & 0.63053                    & 4.44802                    & 0.16          \\ \hline
vits                              & 0.28379                    & 4.34954                    & 0.36          & 0.32154                    & 4.02177                    & 0.36          & 0.30435                    & 4.24128                    & 0.39          \\
vits-ft                           & 0.28692                    & 4.31392                    & 0.36          & 0.33293                    & 3.99321                    & 0.37          & 0.30231                    & 4.21345                    & 0.39          \\
vits-rl                           & \textbf{0.36877}           & \textbf{4.30992}           & \textbf{0.34} & \textbf{0.40277}           & \textbf{4.27409}           & \textbf{0.34} & \textbf{0.36241}           & \textbf{4.31885}           & \textbf{0.35} \\ \hline
diffusion                         & 0.31847                    & 4.32735                    & 0.36          & 0.33100                    & 4.14800                    & 0.36          & 0.32661                    & 4.23370                    & 0.32          \\
diffusion-ft                      & 0.32345                    & 4.23921                    & 0.37          & 0.33313                    & 4.03213                    & 0.36          & 0.32213                    & 4.24315                    & 0.32          \\
diffusion-rl                      & \textbf{0.41125}           & \textbf{4.32996}           & \textbf{0.35} & \textbf{0.41032}           & \textbf{4.32479}           & \textbf{0.35} & \textbf{0.36496}           & \textbf{4.39338}           & \textbf{0.31} \\ \hline
\end{tabular}
\end{table*}

Objective results indicate that our method demonstrates better similarity compared to direct fine-tuning in the SS scenario, while also slightly improving speech quality and intelligibility. MOS scores indicate that the speech quality can be improved by up to 0.25. Similarity scores can be improved by up to 30\% at most.

\subsubsection{Few-Sentence scenario(FS)}

In the FS scenario, speaker representations were extracted from three speeches and averaged to serve as input for the speaker information in both $vits$ and $diffusion$ models. The models, $vits-ft$ and $diffusion-ft$ denote the results after fine-tuning for 2000 steps using three sentences each. Additionally, $vits-rl$ and $diffusion-rl$ represent the fused results achieved through our proposed method.

\begin{table*}[]
\centering
\caption{OBJECTIVE RESULT IN FS SCENARIO}
\label{Obj-ms}
\begin{tabular}{l|ccc|ccc|ccc}
\hline
Objective-FS               & \multicolumn{3}{c|}{libritts-clean}                                     & \multicolumn{3}{c|}{libritts-other}                                     & \multicolumn{3}{c}{vctk}                                                \\ \hline
\multicolumn{1}{c|}{Model} & \multicolumn{1}{c|}{sim ↑} & \multicolumn{1}{c|}{mos ↑} & intell ↓      & \multicolumn{1}{c|}{sim ↑} & \multicolumn{1}{c|}{mos ↑} & intell ↓      & \multicolumn{1}{c|}{sim ↑} & \multicolumn{1}{c|}{mos ↑} & intell ↓      \\ \hline
GT                         & 0.56839                    & 4.52525                    & 0.18          & 0.57535                    & 4.49091                    & 0.21          & 0.71054                    & 4.44977                    & 0.19          \\ \hline
vits                       & 0.29279                    & 4.32537                    & 0.28          & 0.42064                    & 4.05951                    & 0.28          & 0.39705                    & 4.18025                    & 0.29          \\
vits-ft                    & 0.30198                    & 4.24464                    & 0.33          & 0.41559                    & 4.04842                    & 0.27          & 0.24096                    & 4.11934                    & 0.31          \\
vits-rl-3                  & \textbf{0.36375}           & \textbf{4.35064}           & \textbf{0.28} & \textbf{0.44613}           & \textbf{4.29665}           & \textbf{0.27} & \textbf{0.32525}           & \textbf{4.31471}           & \textbf{0.21} \\ \hline
diffusion                  & 0.27494                    & 4.32701                    & 0.41          & 0.29533                    & 4.08498                    & 0.44          & 0.23425                    & 4.24455                    & 0.28          \\
diffusion-ft               & 0.24357                    & 4.23359                    & 0.44          & 0.30260                    & 3.96721                    & 0.43          & 0.40601                    & 4.17133                    & 0.31          \\
diffusion-rl-3             & \textbf{0.32854}           & \textbf{4.31744}           & \textbf{0.36} & \textbf{0.36336}           & \textbf{4.39344}           & \textbf{0.39} & \textbf{0.43316}           & \textbf{4.29905}           & \textbf{0.25} \\ \hline
\end{tabular}
\end{table*}

The experimental results indicate that in the FS scenario, our method achieved the best results in terms of similarity while maintaining stable speech quality and intelligibility. This demonstrates the effectiveness and stability of our proposed approach.

The improvement in similarity can reach up to 50\% at most.

\subsubsection{ASSRL vs. Global TTS model fine-tuning}

In the above experiments, we have demonstrated the effectiveness of ASSRL and shown its superiority over fine-tuning in both SS and FS scenarios.
However, fine-tuning remains the most commonly used method for model adaptation. It is yet to be confirmed whether ASSRL maintains its advantage with an increasing number of reference speeches samples.

In this experiment, we tested the similarity differences between ASSRL and fine-tuning with varying amounts of reference speeches.
The number of speech sentences gradually increases from 2 to 50. Fine-tuning is conducted for 2000 steps each time. When there is only one reference speech sentence, we employ the action strategy in the SS scenario. The current experiment only considers the FS scenario and adopts the corresponding fusion action strategy. The experimental results are presented in Table \ref{Stress}.

\begin{table*}[]
\centering
\caption{ASSRL vs. Global TTS model fine-tuning}
\label{Stress}
\begin{tabular}{l|ccccccc}
\hline
\multicolumn{1}{c|}{}      & \multicolumn{7}{c}{libritts-clean}                                                                                                                                                                                                                               \\ \hline
\multicolumn{1}{c|}{Model} & \multicolumn{1}{c|}{2}                & \multicolumn{1}{c|}{3}                & \multicolumn{1}{c|}{5}                & \multicolumn{1}{c|}{10}               & \multicolumn{1}{c|}{15}               & \multicolumn{1}{c|}{20}               & 50               \\ \hline
vits-ft                    & \multicolumn{1}{c|}{0.28391}          & \multicolumn{1}{c|}{0.28921}          & \multicolumn{1}{c|}{0.28754}          & \multicolumn{1}{c|}{0.29384}          & \multicolumn{1}{c|}{0.30341}          & \multicolumn{1}{c|}{\textbf{0.33135}} & \textbf{0.33299} \\ \hline
vits-rl                    & \multicolumn{1}{c|}{\textbf{0.32839}} & \multicolumn{1}{c|}{\textbf{0.33118}} & \multicolumn{1}{c|}{\textbf{0.33759}} & \multicolumn{1}{c|}{\textbf{0.32479}} & \multicolumn{1}{c|}{\textbf{0.32711}} & \multicolumn{1}{c|}{0.32682}          & 0.32864          \\ \hline
diffusion-ft               & \multicolumn{1}{c|}{0.30291}          & \multicolumn{1}{c|}{0.30402}          & \multicolumn{1}{c|}{0.30659}          & \multicolumn{1}{c|}{0.30814}          & \multicolumn{1}{c|}{\textbf{0.35895}} & \multicolumn{1}{c|}{\textbf{0.32936}} & \textbf{0.34904} \\ \hline
diffusion-rl               & \multicolumn{1}{c|}{\textbf{0.32875}} & \multicolumn{1}{c|}{\textbf{0.33662}} & \multicolumn{1}{c|}{\textbf{0.32758}} & \multicolumn{1}{c|}{\textbf{0.33978}} & \multicolumn{1}{c|}{0.33625}          & \multicolumn{1}{c|}{0.32280}          & 0.32538          \\ \hline
                           & \multicolumn{7}{c}{libritts-other}                                                                                                                                                                                                                               \\ \hline
\multicolumn{1}{c|}{Model} & \multicolumn{1}{c|}{2}                & \multicolumn{1}{c|}{3}                & \multicolumn{1}{c|}{5}                & \multicolumn{1}{c|}{10}               & \multicolumn{1}{c|}{15}               & \multicolumn{1}{c|}{20}               & 50               \\ \hline
vits-ft                    & \multicolumn{1}{c|}{0.31814}          & \multicolumn{1}{c|}{0.30307}          & \multicolumn{1}{c|}{0.30661}          & \multicolumn{1}{c|}{0.31148}          & \multicolumn{1}{c|}{0.33074}          & \multicolumn{1}{c|}{\textbf{0.34058}} & \textbf{0.35146} \\ \hline
vits-rl                    & \multicolumn{1}{c|}{\textbf{0.33797}} & \multicolumn{1}{c|}{\textbf{0.31621}} & \multicolumn{1}{c|}{\textbf{0.31882}} & \multicolumn{1}{c|}{\textbf{0.32031}} & \multicolumn{1}{c|}{\textbf{0.33765}} & \multicolumn{1}{c|}{0.32742}          & 0.32025          \\ \hline
diffusion-ft               & \multicolumn{1}{c|}{0.32803}          & \multicolumn{1}{c|}{0.33019}          & \multicolumn{1}{c|}{0.32784}          & \multicolumn{1}{c|}{0.33104}          & \multicolumn{1}{c|}{0.33566}          & \multicolumn{1}{c|}{\textbf{0.36518}} & \textbf{0.37768} \\ \hline
diffusion-rl               & \multicolumn{1}{c|}{\textbf{0.33499}} & \multicolumn{1}{c|}{\textbf{0.33911}} & \multicolumn{1}{c|}{\textbf{0.33248}} & \multicolumn{1}{c|}{\textbf{0.33026}} & \multicolumn{1}{c|}{\textbf{0.33668}} & \multicolumn{1}{c|}{0.33931}          & 0.32697          \\ \hline
                           & \multicolumn{7}{c}{vctk}                                                                                                                                                                                                                                         \\ \hline
\multicolumn{1}{c|}{Model} & \multicolumn{1}{c|}{2}                & \multicolumn{1}{c|}{3}                & \multicolumn{1}{c|}{5}                & \multicolumn{1}{c|}{10}               & \multicolumn{1}{c|}{15}               & \multicolumn{1}{c|}{20}               & 50               \\ \hline
vits-ft                    & \multicolumn{1}{c|}{0.31847}          & \multicolumn{1}{c|}{0.31529}          & \multicolumn{1}{c|}{0.30739}          & \multicolumn{1}{c|}{0.31859}          & \multicolumn{1}{c|}{0.32195}          & \multicolumn{1}{c|}{\textbf{0.33293}} & \textbf{0.35133} \\ \hline
vits-rl                    & \multicolumn{1}{c|}{\textbf{0.33591}} & \multicolumn{1}{c|}{\textbf{0.32623}} & \multicolumn{1}{c|}{\textbf{0.34254}} & \multicolumn{1}{c|}{\textbf{0.33825}} & \multicolumn{1}{c|}{\textbf{0.34293}} & \multicolumn{1}{c|}{0.32734}          & 0.33547          \\ \hline
diffusion-ft               & \multicolumn{1}{c|}{0.32097}          & \multicolumn{1}{c|}{0.32279}          & \multicolumn{1}{c|}{0.31731}          & \multicolumn{1}{c|}{0.32115}          & \multicolumn{1}{c|}{0.34137}          & \multicolumn{1}{c|}{\textbf{0.36148}} & \textbf{0.36272} \\ \hline
diffusion-rl               & \multicolumn{1}{c|}{\textbf{0.34270}} & \multicolumn{1}{c|}{\textbf{0.35963}} & \multicolumn{1}{c|}{\textbf{0.36897}} & \multicolumn{1}{c|}{\textbf{0.35302}} & \multicolumn{1}{c|}{\textbf{0.35353}} & \multicolumn{1}{c|}{0.34337}          & 0.35879          \\ \hline
\end{tabular}
\end{table*}

The experimental results indicate that ASSRL performs better than fine-tuning when the number of reference speeches is fewer than 20. However, once this threshold is surpassed, fine-tuning begins to show its advantage. This suggests that fine-tuning the model requires a sufficient amount of data to effectively prevent overfitting. Furthermore, increasing the number of reference speech sentences from 5 to 10 does not lead to a significant improvement in similarity. Therefore, in the FS scenario, using just 3 sentences is sufficient and also facilitates data collection.

\begin{table*}[]
\centering
\caption{Fusion scoring mechanism ablation results(Bold top 50\%)}
\label{scoreablation}
\resizebox{\textwidth}{!}{%
\begin{tabular}{|cccccccccccc|}
\hline
\multicolumn{12}{|c|}{\textit{vits-rl(SS)}}                                                                                                                                                                                                                                                                                                                                                                                                                                                       \\ \hline
\multicolumn{1}{|c|}{\multirow{2}{*}{Similarity}} & \multicolumn{1}{c|}{\multirow{2}{*}{Mos}} & \multicolumn{1}{c|}{\multirow{2}{*}{Intelligibility}} & \multicolumn{3}{c|}{Similarity ↑}                                                                                     & \multicolumn{3}{c|}{Mos ↑}                                                                                            & \multicolumn{3}{c|}{Intelligibility ↓}                                                    \\ \cline{4-12} 
\multicolumn{1}{|c|}{}                            & \multicolumn{1}{c|}{}                     & \multicolumn{1}{c|}{}                                 & \multicolumn{1}{c|}{libritts-clean}   & \multicolumn{1}{c|}{libritts-other}   & \multicolumn{1}{c|}{vctk}             & \multicolumn{1}{c|}{libritts-clean}   & \multicolumn{1}{c|}{libritts-other}   & \multicolumn{1}{c|}{vctk}             & \multicolumn{1}{c|}{libritts-clean} & \multicolumn{1}{c|}{libritts-other} & vctk          \\ \hline
\multicolumn{1}{|c|}{\multirow{4}{*}{\checkmark}} & \multicolumn{1}{c|}{\checkmark}           & \multicolumn{1}{c|}{\checkmark}                       & \multicolumn{1}{c|}{\textbf{0.28223}} & \multicolumn{1}{c|}{0.30808}          & \multicolumn{1}{c|}{\textbf{0.29414}} & \multicolumn{1}{c|}{\textbf{4.44522}} & \multicolumn{1}{c|}{\textbf{4.27828}} & \multicolumn{1}{c|}{\textbf{4.36816}} & \multicolumn{1}{c|}{\textbf{0.28}}  & \multicolumn{1}{c|}{\textbf{0.30}}  & \textbf{0.24} \\
\hline
\end{tabular}%
}
\end{table*}

\begin{table*}[]
\centering
\caption{State Representation Ablation Results(Bold Top 25\% )}
\label{stateablation}
\resizebox{\textwidth}{!}{%
\begin{tabular}{|cccccccccccccc|}
\hline
\multicolumn{14}{|c|}{\textit{vits-rl (SS)}}                                                                                                                                                                                                                                                                                                                                                                                                                                                                                                                             \\ \hline
\multicolumn{3}{|c|}{prior}                                                                                                                              & \multicolumn{2}{c|}{post}                                         & \multicolumn{3}{c|}{Similarity ↑}                                                                                     & \multicolumn{3}{c|}{mos ↑}                                                                                            & \multicolumn{3}{c|}{Intelligibility ↓}                                                    \\ \hline
\multicolumn{1}{|c|}{$e$}                          & \multicolumn{1}{c|}{$f_{rv}$}                    & \multicolumn{1}{c|}{$f_t$}                       & \multicolumn{1}{c|}{$e_s$}      & \multicolumn{1}{c|}{$f_{sv}$}   & \multicolumn{1}{c|}{libritts-clean}   & \multicolumn{1}{c|}{libritts-other}   & \multicolumn{1}{c|}{vctk}             & \multicolumn{1}{c|}{libritts-clean}   & \multicolumn{1}{c|}{libritts-other}   & \multicolumn{1}{c|}{vctk}             & \multicolumn{1}{c|}{libritts-clean} & \multicolumn{1}{c|}{libritts-other} & vctk          \\
\hline
\end{tabular}%
}
\end{table*}

\subsection{Ablation Experiment}
In this section, training time is set be consistent, with each model learning the same data under each configuration. Similarity scores are kept constant in the fusion scoring mechanism, and speaker representation are maintained in the prior state design. Ablation experiments are conducted under these conditions.
Experimental results are objectively evaluated based on similarity, speech quality, and speech intelligibility. Results are presented separately for SS and FS scenarios.

\subsubsection{Fusion scoring mechanism}
Through ablation experiments, the effectiveness of the fusion scoring mechanism was validated in both SS and FS scenarios. In the FS scenario, the number of reference speeches was set to 3. Apart from similarity scoring, each model was evaluated across four scenarios for each dataset: preserving both speech quality and intelligibility, preserving only one, and preserving neither. 
The results are presented in Table \ref{scoreablation}.

The experimental results indicate that while focusing solely on similarity scores improved performance in both scenarios, there was a noticeable decline in speech quality and intelligibility. This highlights the efficacy of the fusion scoring mechanism, which penalizes decreases in quality and intelligibility. This prevents reinforcement learning from overly prioritizing similarity improvements during exploration, thereby mitigating the risk of selecting incorrect actions.

\subsubsection{State representation}
We verify the rationality of the state representation design through ablation experiments. Our objective is to enhance speaker similarity, focusing on directly related feature information such as speaker embeddings and voiceprint features. Additionally, since synthesizing speech with the same speaker embedding but different texts yields varying similarity outcomes, we also consider text features. 
We categorize the speaker embedding \( e \) and the voiceprint feature \( f_{rv} \) obtained from the reference speech, along with the feature \( f_t \) derived from the text, as prior knowledge. The speaker embedding \( e_s \) and the voiceprint feature \( f_{sv} \) extracted from the synthesized speech are classified as posterior knowledge. Given our goal of improving speaker embedding \( e \), this feature remains constant. We conduct separate tests on the effects of these features in both SS and FS scenarios. 
In the FS scenario, where the number of reference speeches is 3, the voiceprint feature \( f_{rv} \) is calculated as the average of all voiceprint features from the reference speeches. Initially, the speaker embedding \( e \) is also computed as the average of corresponding features from the reference speeches.
Additionally, we highlight the top 25\% of results for each model within a specific dataset.
The results are presented in Table \ref{stateablation}.

After observing the experimental results, we found that including only the speaker embedding \( e \) and the text feature \( f_t \) in the state representation yields the best overall performance. This configuration achieves high similarity while maintaining stable speech quality and intelligibility.
Including all features achieves the highest similarity among all experimental setups but may result in decreased speech quality. Maintaining consistency in prior information reveals that as more posterior information is provided, speech quality tends to decrease.
When only \( f_{rv} \) is included as prior information, the improvement in similarity is relatively small. Therefore, we propose representing the state comprehensively as \( s = (f_t, e) \).

\subsection{Other experiments}

In this section, we mainly explore the use of pre-trained voiceprint analysis models and the setting of some hyperparameters.

\subsubsection{Voiceprint analysis model}

We evaluated three open-source speaker verification models with different structures \cite{DBLP:conf/interspeech/DesplanquesTD20,X-vectors,VILLALBA2020101026}. From each dataset's training, validation, and test sets, we randomly selected 10 speakers, and for each speaker, we randomly selected 10 speech samples. The voiceprint analysis models were then used to determine whether the speech samples were from the same speaker, and the accuracy of the models was calculated. The results are shown in Table \ref{voiceprintmodel}.

\begin{table*}[]
\centering
\caption{Accuracy of Voiceprint Analysis Models}
\label{voiceprintmodel}
\begin{tabular}{l|lll|lll|lll}
\hline
\multirow{2}{*}{Model} & \multicolumn{3}{c|}{libritts-clean}                                                                   & \multicolumn{3}{c|}{libritts-other}                                                                   & \multicolumn{3}{c}{vctk}                                                                             \\ \cline{2-10} 
                       & \multicolumn{1}{c|}{train}          & \multicolumn{1}{c|}{dev}            & \multicolumn{1}{c|}{test} & \multicolumn{1}{c|}{train}          & \multicolumn{1}{c|}{dev}            & \multicolumn{1}{c|}{test} & \multicolumn{1}{c|}{train}          & \multicolumn{1}{c|}{dev}            & \multicolumn{1}{c}{test} \\ \hline
Ecapa                  & \multicolumn{1}{l|}{0.732}          & \multicolumn{1}{l|}{0.897}          & 0.835                     & \multicolumn{1}{l|}{0.817}          & \multicolumn{1}{l|}{0.856}          & 0.661                     & \multicolumn{1}{l|}{0.795}          & \multicolumn{1}{l|}{0.821}          & 0.851                    \\ \hline
X-Vector               & \multicolumn{1}{l|}{\textbf{0.886}} & \multicolumn{1}{l|}{\textbf{0.991}} & \textbf{0.986}            & \multicolumn{1}{l|}{\textbf{0.985}} & \multicolumn{1}{l|}{\textbf{0.878}} & \textbf{0.978}            & \multicolumn{1}{l|}{\textbf{0.945}} & \multicolumn{1}{l|}{\textbf{0.887}} & \textbf{0.872}           \\ \hline
ResNet                 & \multicolumn{1}{l|}{0.757}          & \multicolumn{1}{l|}{0.814}          & 0.852                     & \multicolumn{1}{l|}{0.885}          & \multicolumn{1}{l|}{0.827}          & 0.681                     & \multicolumn{1}{l|}{0.870}          & \multicolumn{1}{l|}{0.767}          & 0.797                    \\ \hline
\end{tabular}
\end{table*}

\begin{table*}[]
\centering
\caption{The settings for $\lambda_1$ and $\lambda_2$}
\label{lamda}
\begin{tabular}{c|c|ccc|ccc|ccc|ccc}
\hline
\multirow{2}{*}{Model}   & \multirow{2}{*}{Dataset} & \multicolumn{3}{c|}{$\lambda_1$=1,$\lambda_2$=1}                                     & \multicolumn{3}{c|}{$\lambda_1$=0.5,$\lambda_2$=0.5}                        & \multicolumn{3}{c|}{$\lambda_1$=0.1,$\lambda_2$=0.1}                            & \multicolumn{3}{c}{$\lambda_1$=0.5,$\lambda_2$=0.1}                         \\ \cline{3-14} 
                         &                          & \multicolumn{1}{c|}{Sim ↑}   & \multicolumn{1}{c|}{Mos ↑}            & Intell ↓      & \multicolumn{1}{c|}{Sim ↑}   & \multicolumn{1}{c|}{Mos ↑}   & Intell ↓      & \multicolumn{1}{c|}{Sim↑}             & \multicolumn{1}{c|}{Mos ↑}   & Intell ↓ & \multicolumn{1}{c|}{Sim ↑}   & \multicolumn{1}{c|}{Mos ↑}   & Intell ↓      \\ \hline
\multirow{3}{*}{vits-rl} & libritts-clean           & \multicolumn{1}{c|}{0.20392} & \multicolumn{1}{c|}{\textbf{4.43912}} & \textbf{0.19} & \multicolumn{1}{c|}{0.24832} & \multicolumn{1}{c|}{4.39384} & \textbf{0.19} & \multicolumn{1}{c|}{\textbf{0.30291}} & \multicolumn{1}{c|}{4.03112} & 0.22     & \multicolumn{1}{c|}{0.30023} & \multicolumn{1}{c|}{4.30281} & \textbf{0.19} \\ \cline{2-14} 
                         & libritts-other           & \multicolumn{1}{c|}{0.19232} & \multicolumn{1}{c|}{\textbf{4.29314}} & \textbf{0.18} & \multicolumn{1}{c|}{0.23481} & \multicolumn{1}{c|}{4.19312} & 0.19          & \multicolumn{1}{c|}{\textbf{0.29381}} & \multicolumn{1}{c|}{3.93811} & 0.23     & \multicolumn{1}{c|}{0.28379} & \multicolumn{1}{c|}{4.28192} & 0.21          \\ \cline{2-14} 
                         & vctk                     & \multicolumn{1}{c|}{0.21832} & \multicolumn{1}{c|}{\textbf{4.39123}} & \textbf{0.27} & \multicolumn{1}{c|}{0.25378} & \multicolumn{1}{c|}{4.29314} & 0.28          & \multicolumn{1}{c|}{\textbf{0.29482}} & \multicolumn{1}{c|}{4.03914} & 0.31     & \multicolumn{1}{c|}{0.28281} & \multicolumn{1}{c|}{4.29329} & 0.28          \\ \hline
\multirow{3}{*}{diff-rl} & libritts-clean           & \multicolumn{1}{c|}{0.20510} & \multicolumn{1}{c|}{\textbf{4.40462}} & \textbf{0.28} & \multicolumn{1}{c|}{0.25521} & \multicolumn{1}{c|}{4.36542} & 0.31          & \multicolumn{1}{c|}{\textbf{0.28833}} & \multicolumn{1}{c|}{4.16247} & 0.32     & \multicolumn{1}{c|}{0.27035} & \multicolumn{1}{c|}{4.23071} & 0.31          \\ \cline{2-14} 
                         & libritts-other           & \multicolumn{1}{c|}{0.21293} & \multicolumn{1}{c|}{\textbf{4.20312}} & \textbf{0.31} & \multicolumn{1}{c|}{0.26432} & \multicolumn{1}{c|}{4.18292} & 0.31          & \multicolumn{1}{c|}{\textbf{0.30892}} & \multicolumn{1}{c|}{3.92813} & 0.33     & \multicolumn{1}{c|}{0.30102} & \multicolumn{1}{c|}{4.19823} & 0.31          \\ \cline{2-14} 
                         & vctk                     & \multicolumn{1}{c|}{0.22583} & \multicolumn{1}{c|}{\textbf{4.38821}} & \textbf{0.28} & \multicolumn{1}{c|}{0.24831} & \multicolumn{1}{c|}{4.31242} & 0.28          & \multicolumn{1}{c|}{\textbf{0.31891}} & \multicolumn{1}{c|}{3.93842} & 0.29     & \multicolumn{1}{c|}{0.29333} & \multicolumn{1}{c|}{4.29311} & 0.28          \\ \hline
\end{tabular}
\end{table*}

After evaluation, we found that the voiceprint analysis model based on the X-Vector structure has a very high accuracy, so we believe that the similarity score given by the model is the most reliable. Based on the experimental results, we chose the X-Vector voiceprint analysis model as the similarity score in the fusion scoring mechanism.

\begin{figure*}[!t]
	\centering
	
	\subfloat[gamma-vits]{\includegraphics[width=0.25\linewidth]{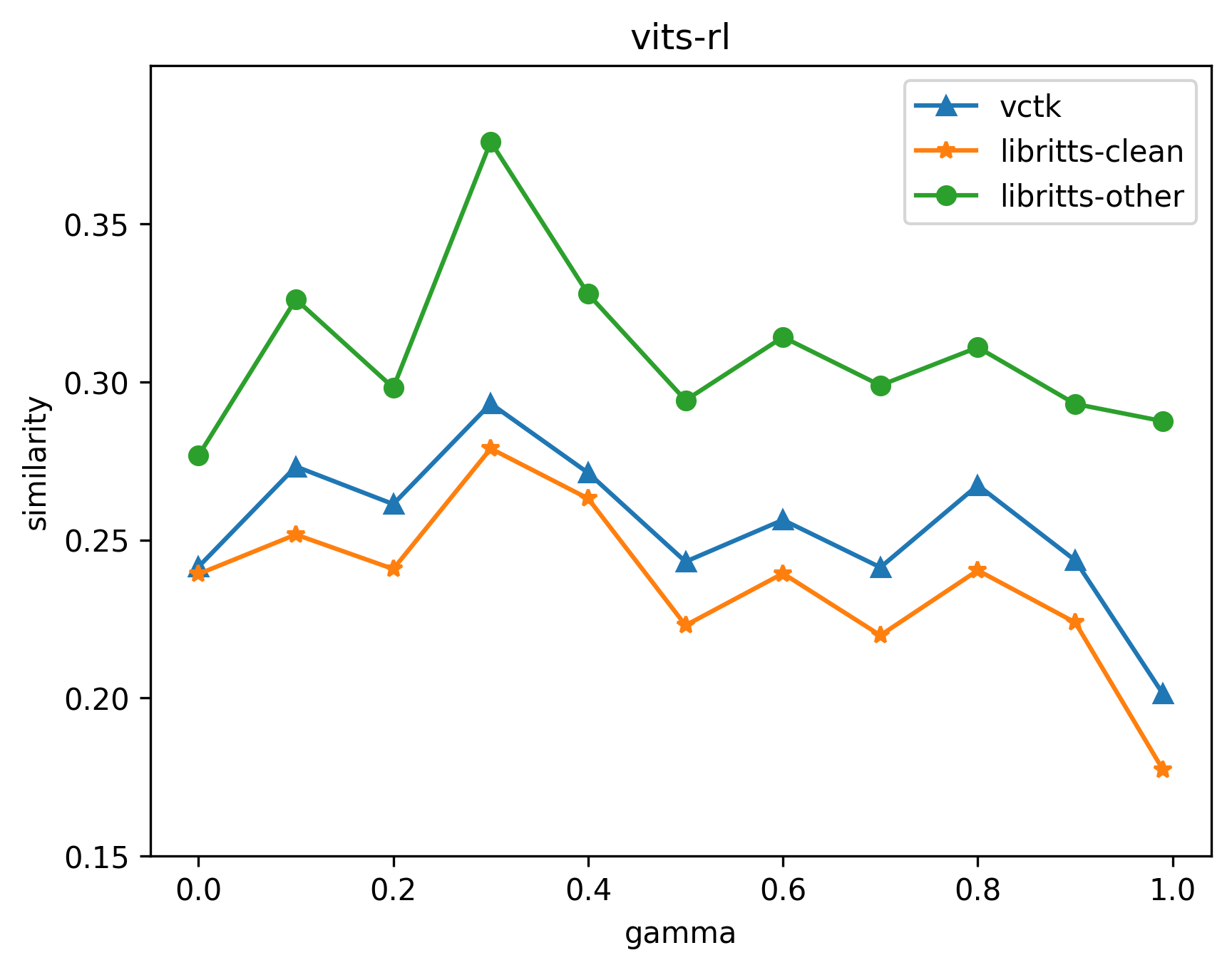}%
		\label{fig_first_case}}
	\subfloat[gamma-diffusion]{\includegraphics[width=0.25\linewidth]{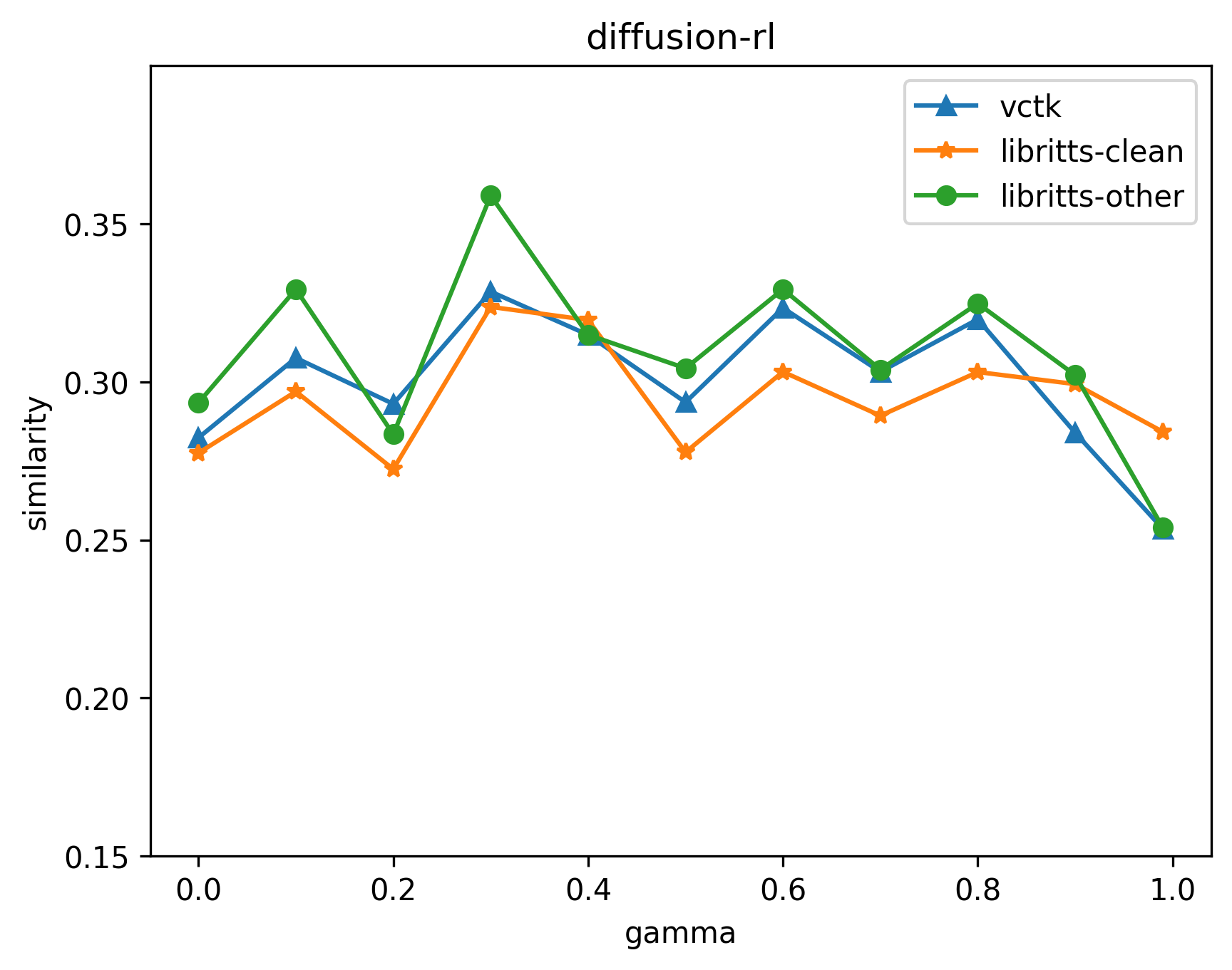}
		\label{fig_first_case}}
	\subfloat[action factor-vits]{\includegraphics[width=0.25\linewidth]{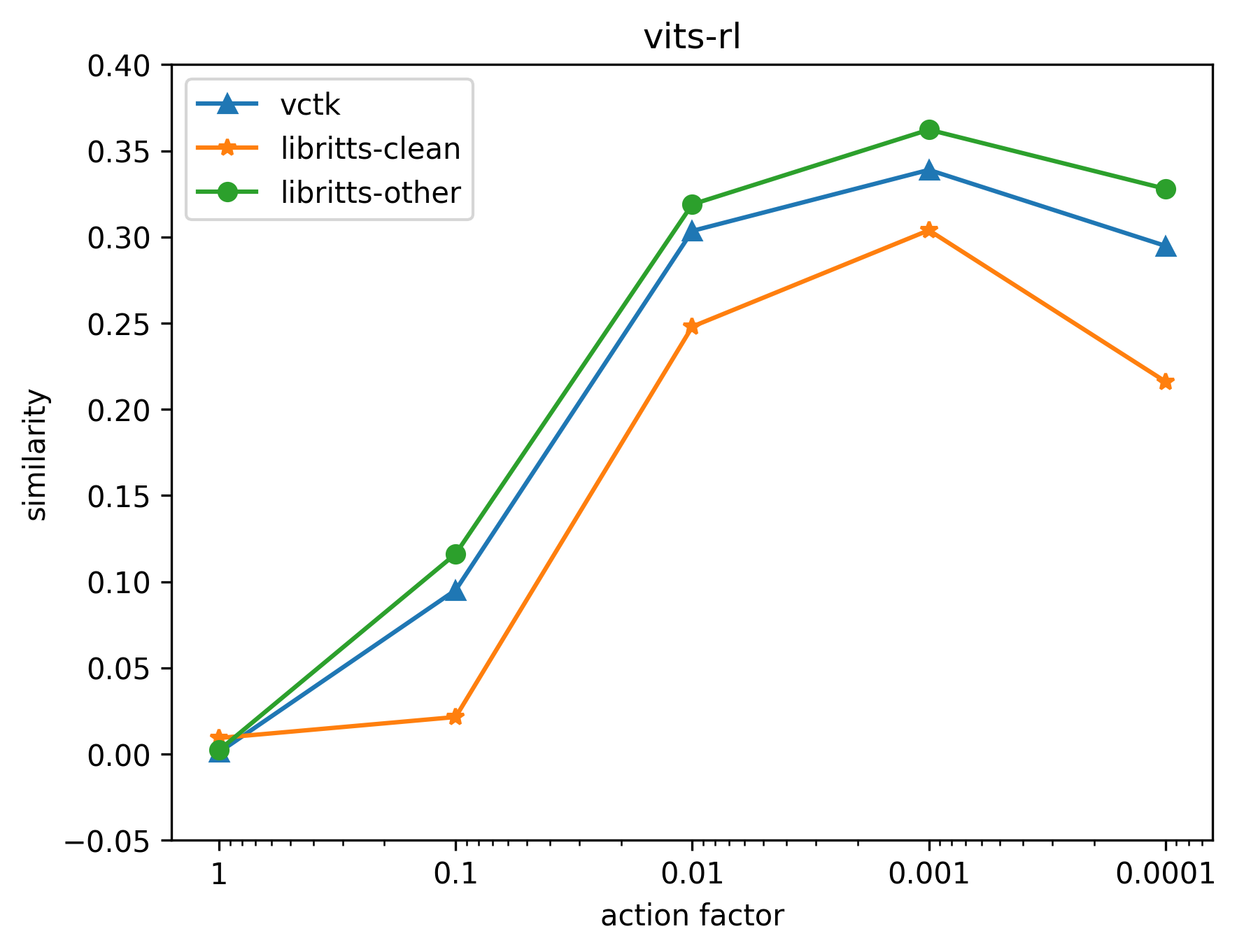}%
		\label{fig_first_case}}
	\subfloat[action factor-diffusion]{\includegraphics[width=0.25\linewidth]{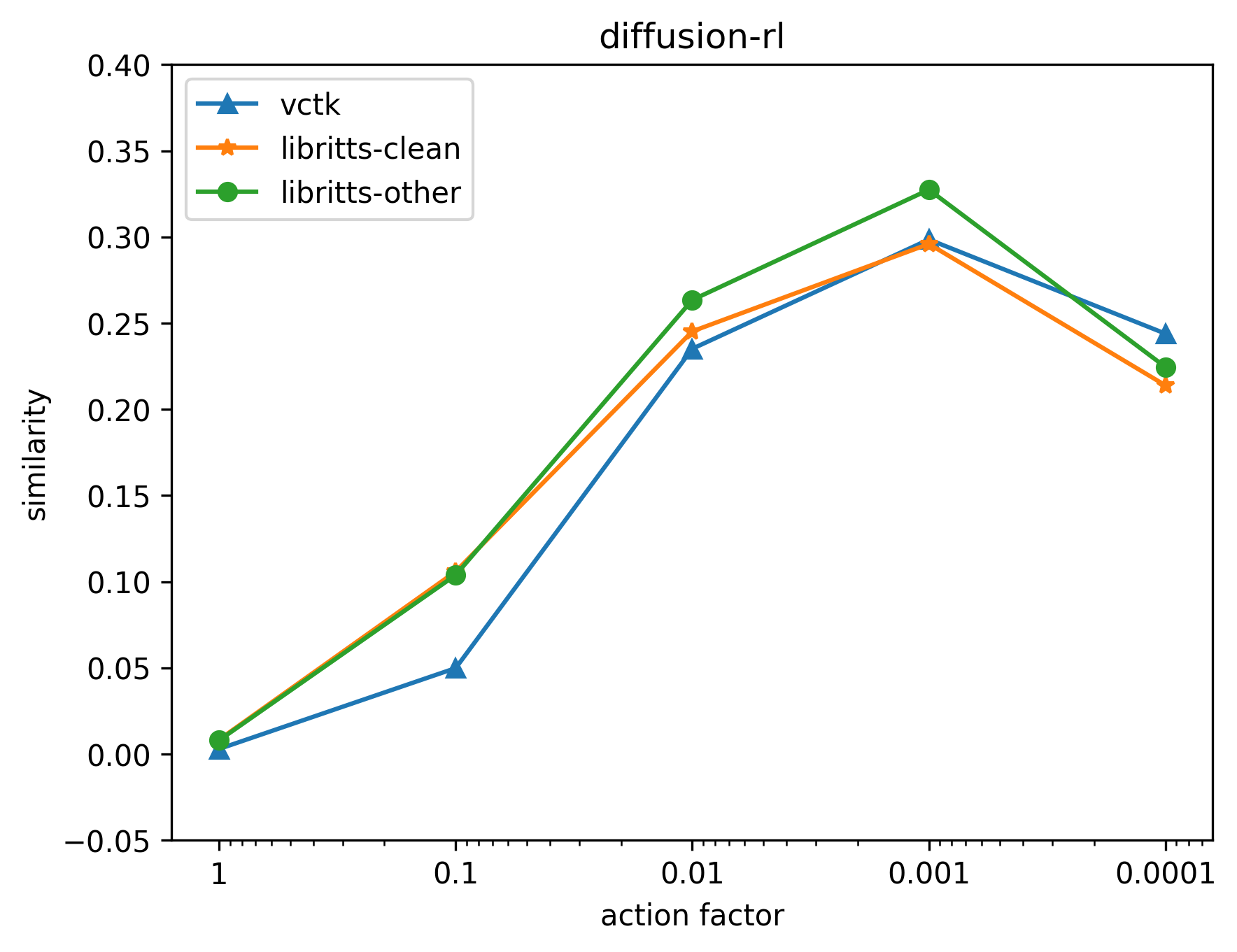}%
		\label{fig_first_case}}

	\subfloat[step-vits]{\includegraphics[width=0.25\linewidth]{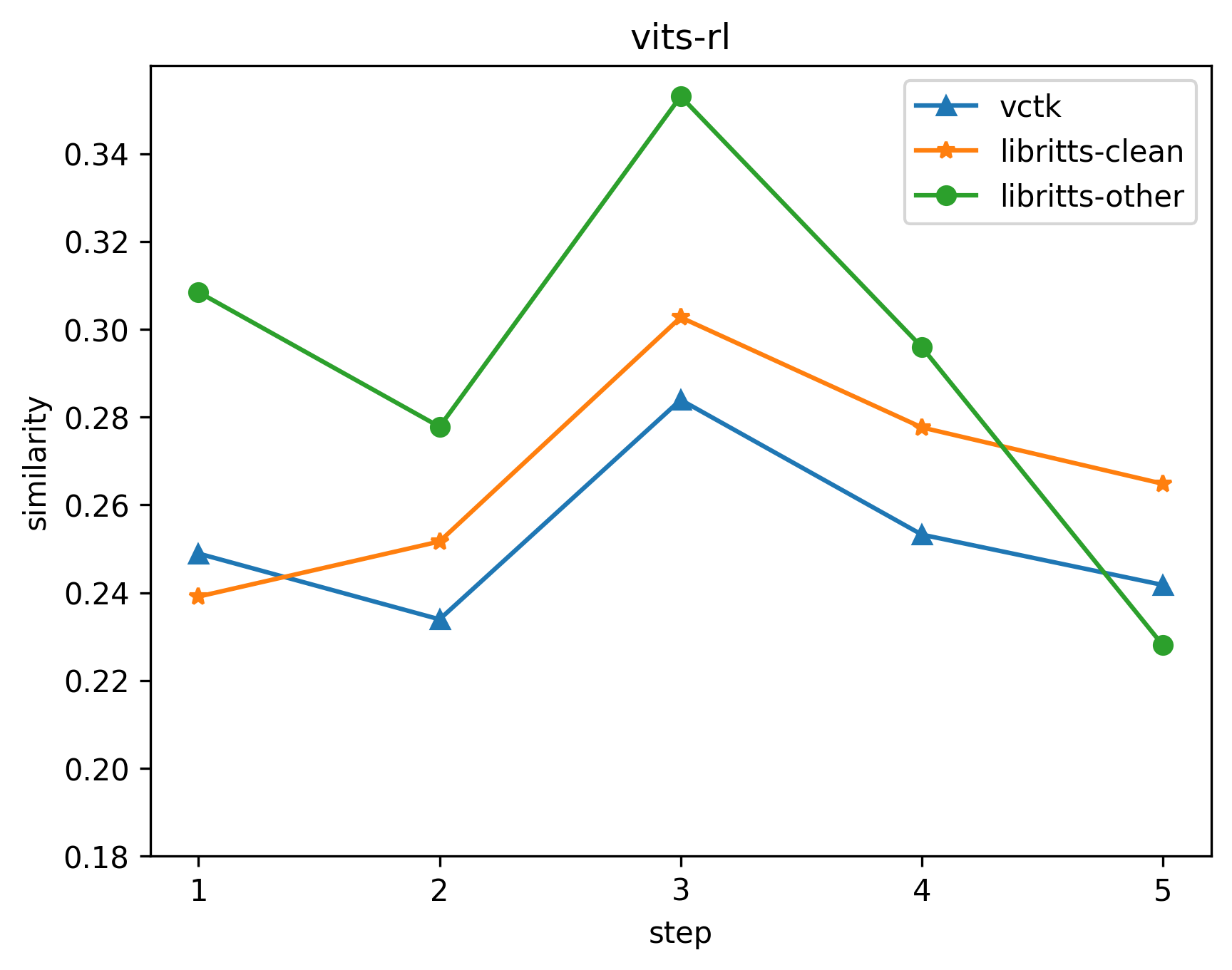}%
		\label{fig_first_case}}
	\subfloat[step-diffusion]{\includegraphics[width=0.25\linewidth]{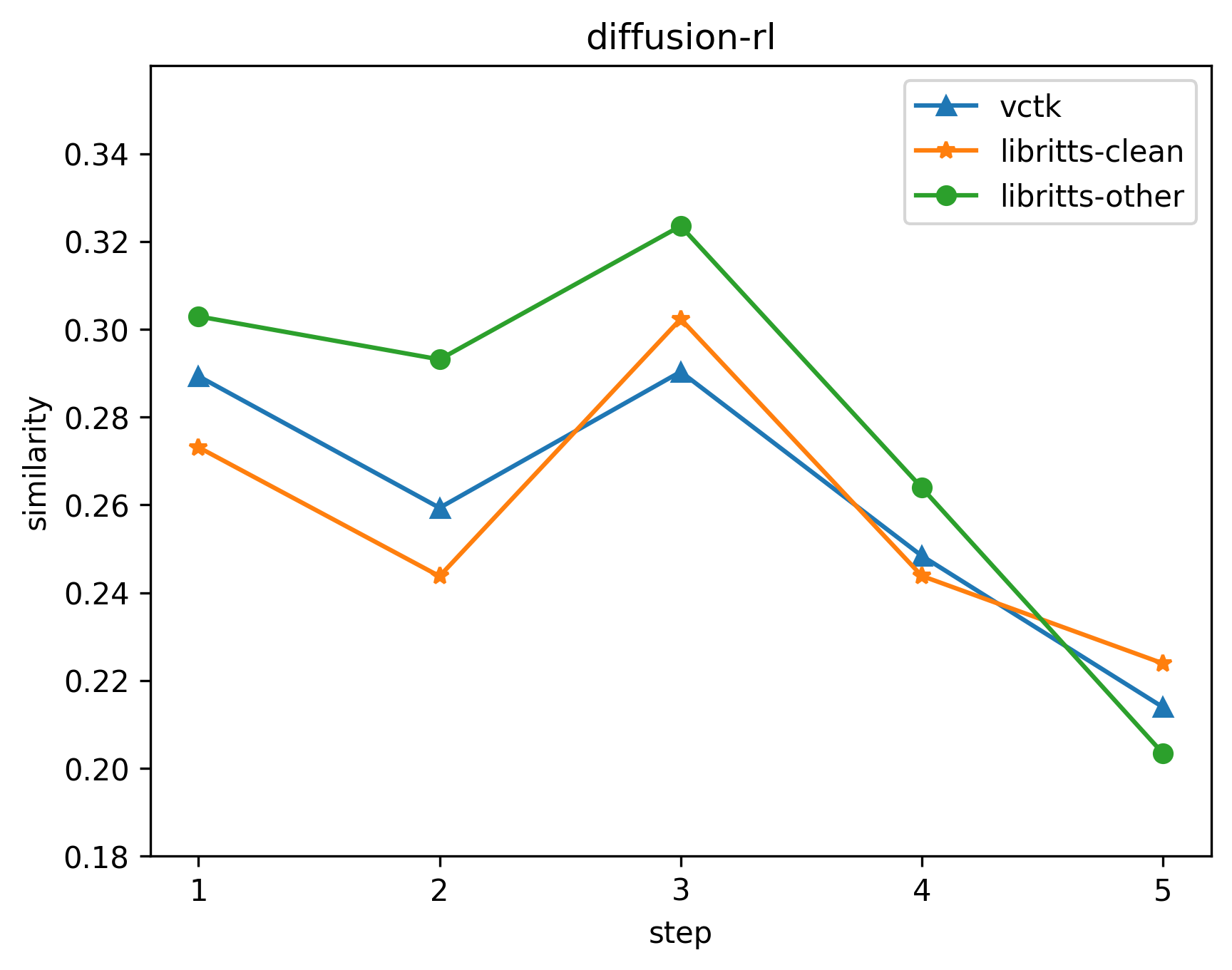}%
		\label{fig_first_case}}
	\caption{The results of hyperparameter changes in reinforcement learning. (a) and (b) show the impact of changes in gamma values on similarity during reinforcement learning. (c) and (d) show the impact of changes in action scaling factor. (e) and (f) show the impact of changes in the number of exploration steps.}
	\label{hyper}
\end{figure*}

\subsubsection{Hyperparameter settings}

In this section, we primarily explore the settings for the values of \(\lambda_1\) and \(\lambda_2\) in formula (1), as well as the settings for the gamma value in RL, the scaling factor for actions, and the number of exploration steps through experiments. These experiments were conducted in the SS scenario, and their results were directly applied to the FS scenario.

The settings for $\lambda$ are shown in Table \ref{lamda}. After testing each model on each dataset with various configurations, we found that excessively large values, while ensuring high speech quality and intelligibility, reduce the importance of speech similarity, leading to a decrease in its results. Conversely, setting the values too low would cause the policy network to deceive the reward model in pursuit of higher similarity, ultimately resulting in poor speech quality. 
Based on the experimental results, setting \(\lambda_1\) to 0.5 and \(\lambda_2\) to 0.1 can achieve a better balance among multiple scores.

The experimental results for RL hyperparameters are shown in Figure \ref{hyper}.
We explored the impact of these parameters on similarity by varying the relevant parameters.

During the experiments on the impact of changes in the gamma value, the action scaling factor was set to 0.001 and the exploration steps were set to 2.
(a) and (b) respectively show the performance improvement in speaker similarity for vits and diffusion models on different datasets using our method as gamma increases from 0 to 0.99.
Both models exhibit high sensitivity to changes in gamma values, with the similarity result being optimal when the gamma value is set to 0.3.

(c) and (d) show the changes in speaker similarity resulting from variations in the action scaling factor. Each experiment was trained for the same amount of time, with the gamma value set to 0.99 and the exploration steps set to 2.
The experimental results indicate that the best performance is achieved with a 1000-fold scaling.
Theoretically, with sufficient training time, the results with a scaling factor of 0.0001 could be further improved, but this would require more computational time and resources.
Additionally, we speculate that the poor results with a too-small scaling factor occur because, although the speaker embedding \(e\) has room for improvement, the distance between the improved embedding \(e^*\) and \(e\) won't be too large. If the adjustment granularity is too coarse, it can lead to an excessively large distance. Once this distance exceeds a reasonable range, it becomes difficult for subsequent exploration steps to quickly return to the starting point, making it challenging to ultimately find the optimal result.
Based on the experimental results, we set this parameter to 0.001.

(e) and (f) show the changes in speech similarity as the exploration steps increase from 1 to 5. Each experiment was trained for the same amount of time, with the gamma value set to 0.99 and the action scaling factor set to 0.001.
The experimental results indicate that increasing the exploration steps to 3 leads to higher similarity. However, continuing to increase the exploration steps results in a decrease in similarity.
We believe that the number of steps is also constrained by the distance between \(e\) and \(e^*\).
In limited training time, excessive exploration steps can easily lead to stepping out of this range, resulting in a decrease in similarity.

\section{Conclusion}
In this paper, we propose an Agile Speaker Representation Reinforcement Learning (ASRRL) strategy for the TTS speaker adaptation task. The ASRRL strategy leverages prior knowledge to enhance the similarity of synthetic speech in the target domain. The state representation for reinforcement learning is meticulously designed to meet the task requirements, and different action strategies are proposed for Single-Sentence(SS) and Few-Sentence(FS) scenarios. In the SS scenario, we introduce an accelerated retrieval of speaker refinement items. For the FS scenario, we propose a dynamic fusion action strategy to adaptively integrate a few reference speeches. Additionally, a fusion scoring mechanism is designed to comprehensively evaluate the exploration process across three dimensions. Subjective and objective evaluations demonstrate that the proposed ASRRL method ensures stability in naturalness and intelligibility while enhancing speaker similarity. Objective evaluations show that when the number of reference speech sentences is below 20, the ASRRL method outperforms the fine-tuning strategy. Ablation experiments indicate that multidimensional integration in the scoring mechanism is necessary to prevent sub-optimal results in a single dimension. Adding text features to the state representation enhances the reinforcement learning agent's ability to understand the target and output more reasonable actions.
Finally, we investigate the hyperparameters of RL through a series of experiments and set reasonable values based on the results. In future work, we plan to apply reinforcement learning strategies to a broader range of audio generation tasks.

\bibliographystyle{IEEEtran}\
\bibliography{zerotts}

\newpage

\vfill

\end{document}